\documentclass[epj,nopacs]{svjour}
\usepackage{graphics}

\begin{document}

\title{Resonant antineutrino induced electron capture with low energy bound-beta beams} 

\author{R. G. C. Oldeman \and M. Meloni \and B. Saitta}
\institute{Dipartimento di fisica, Universit\`a degli Studi di Cagliari and INFN, Sezione di Cagliari}

\date{Received: date / Revised version: date}

\abstract{
Antineutrino induced electron capture is a resonant process that can have a large cross-section
for beams of monochromatic antineutrinos.
We calculate the cross-section of this process and investigate
an experimental setup where monochromatic antineutrinos are produced from 
the bound-beta decay of fully ionized radioactive atoms in a storage ring.
If the energy between the source and the target is well matched,
the cross-sections can be significantly
larger than the cross-sections of commonly used non-resonant processes.
The rate that can be achieved at a small distance
between the source and two targets of $10^3$\,kg is up to one interaction per $8.3\cdot10^{18}$ 
decaying atoms.
For a source-target distance corresponding to the first atmospheric neutrino oscillation maximum,
the largest rate is one interaction per $3.2\cdot10^{21}$ decaying atoms,
provided that extremely stringent monochromaticity conditions ($10^{-7}$
or better) are achieved in future ion beams.}

\maketitle

\section{Introduction}
\label{intro}
In ordinary electron capture, an electron from one of the $s$ orbitals is 
absorbed by the nucleus, and a neutrino is emitted.
If this process is energetically forbidden, it can be 
induced by an incoming antineutrino, 
a process called antineutrino induced electron capture ($\bar\nu$EC)~\cite{mikaelyan}.
The cross-section is resonant,
and with a monochromatic source of antineutrinos of the right energy, large cross-sections can be obtained.
The case of the $^3$H\,$\leftrightarrow$\,$^3$He system is of particular interest
since recoilless emission and absorption can result in a 
Mossbauer effect for neutrinos~\cite{visscher,Kells:1984nm,Raghavan:2005gn,Raghavan:2006xf}.
Here we study the process for a large number of target nuclei,
and propose to use bound-beta decays of  fully ionized radioactive atoms 
as a source of monochromatic antineutrinos.
Throughout the paper, we consider only nuclear allowed transitions, 
where the parity of the initial and final state nucleus are equal and the spins differ by at most one unit of $\hbar$.
We indicate two 
possible experimental signatures: to observe the photons from the atomic de-excitation and 
the decay of the daughter nucleus.

\section{Reaction cross-section of antineutrino induced electron capture}
\label{crosssect}

Antineutrino induced electron capture on a neutral atom can be described as:
\begin{equation}\label{eq:nec}
\bar\nu_e+^A_{Z+1}Y\to^A_ZX^*,
\end{equation}
where $^A_ZX$ represents a nucleus with $Z$ protons and $A$ nucleons.
The asterix on the final state atom indicates that the atom is in an excited state:
after one of the inner electrons has been captured, an electron from a higher shell
will fill the vacancy by emitting one or more photons.
Since antineutrino induced electron capture is a two-body process with an unstable intermediate state, 
its cross-section has a resonant character, being largest when the center-of-mass energy $Q$ is equal to
the $Q$ value of the target process $Q_t$, and rapidly decreasing for other energies:
\begin{equation}\label{eq:crs}
\sigma_{\bar\nu{\rm EC}}(Q)=S\frac{4\pi}{Q^2}
\left[\frac{\Gamma^2/4}{(Q-Q_t)^2+\Gamma^2/4}\right]\frac{\Gamma_{b\beta}}{\Gamma},
\end{equation}
where $S=\frac{2J+1}{(2S_1+1)(2S_2+1)}$, a factor depending on the spin $J$ of the resonance
and $S_1$ and $S_2$, the spins of the two initial state particles. 
$S$ is a factor of order 1, and we use the approximation $S=1$.
$\Gamma$ is the total width of the excited atom, and $\Gamma_{b\beta}$ is the 
width of the excited atom to decay back to its original state, through bound-beta decay.
The calculation of $\Gamma_{b\beta}$ is described in Appendix~\ref{sec:appendix}.


The total width $\Gamma$ of an atom with a vacancy of an electron in one of the inner
atomic shells is dominated by radiative de-excitation.
The most likely de-excitation of an atom with an electron vacancy in an $s$ shell 
with a quantum number of energy $n$ is 
an electric dipole transition from an electron in a $(n+1)p$ shell.
We approximate this width as:
\begin{equation}\label{eq:gammarad}
\Gamma=\alpha\frac{B^2}{m_e},
\end{equation}
where $\alpha$ is the fine structure constant, $m_e$ the electron mass, and
$B$ the binding energy of the captured electron (defined positive)~\cite{Ebind}.


The $Q$ value of antineutrino induced electron capture on a neutral atom is:
\begin{equation}\label{eq:Qnec}
Q_t=m(^A_ZX)-m(^A_{Z+1}Y)+B,
\end{equation}
where $m(^A_ZX)$ refers to the mass of the neutral atom~\cite{Audi:2002rp}.
The antineutrino energy must be slightly higher than the $Q$ value, since 
a small fraction of the neutrino energy  is used for the kinetic energy of the final state:
\begin{equation}
E_{\bar\nu t}=Q_t+\frac{Q_t^2}{2m}.
\end{equation}

\begin{figure}
\resizebox{0.50\textwidth}{!}{
  \includegraphics{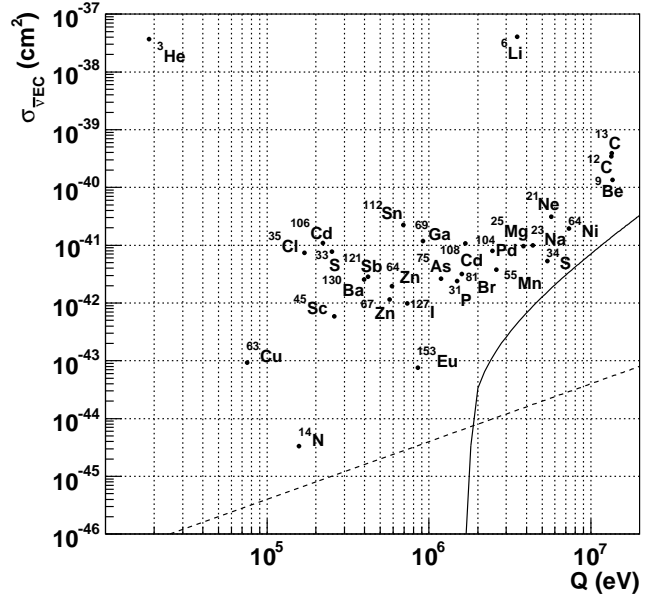}
}
\caption{Illustration of the peak cross-sections of 
antineutrino induced electron capture from the $1s$ orbital for some representative atoms.
For comparison, the solid curve indicates the antineutrino-proton cross-section
and the dashed curve represents the antineutrino-electron cross-section.}
\label{fig:crsplot}
\end{figure}

All  stable atoms, with halflife $t_{1/2}>10^{16}$\,s, and a peak cross-section 
$\sigma_{\rm peak}\geq5\cdot10^{-42}$\,cm$^2$ for
antineutrino induced electron capture from the $1s$ orbital are listed in Tab.~\ref{tab:crs}. 
In Fig.~\ref{fig:crsplot} the peak cross-section for some elements are compared to the 
antineutrino-proton cross-section and the antineutrino-electron cross-section.

Cross-sections for electron capture from higher orbitals are calculated in the same way.
While $\Gamma_{b\beta}$ decreases as $n^{-3}$, the radiative width decreases as $n^{-4}$,
and the peak cross-section scales linearly with $n$.

We find that the peak cross-section for antineutrino induced electron capture
can be several orders of magnitude larger than the commonly used processes
of scattering on protons or electrons.
For example, four targets ($^3$He, $^{69}$Ga, $^{106}$Cd, and $^{112}$Sn) have peak cross-sections
in excess of $10^{-41}$\,cm$^2$ for sub-MeV antineutrinos, while the antineutrino-electron 
cross-section is less than $10^{-44}$\,cm$^2$ at those energies.
 However, the cross-section 
rapidly decreases outside a narrow energy window, and to exploit these large cross-sections
requires a monochromatic source of antineutrinos.

\begin{table}
\caption{List of stable ($t_{1/2}\geq10^{16}$\,s) elements with a peak cross-section for 
antineutrino induced electron capture from the $1s$ orbital of $5\cdot10^{-42}$\,cm$^2$ or larger.}
\begin{center}
\begin{tabular}{ccccc}
\hline\noalign{\smallskip}
   & $Q$    & $\sigma_{\rm peak}$  &  $\Gamma$ & natural\\
   &  (eV) &    (cm$^2$) &   (eV)  &  abundance \\
\noalign{\smallskip}\hline\noalign{\smallskip}
$^{3}$He & $1.86\cdot10^{4}$ & $3.66\cdot10^{-38}$ & $2.64\cdot10^{-6}$ & $1.37\cdot10^{-6}$ \\
$^{6}$Li & $3.51\cdot10^{6}$ & $4.04\cdot10^{-38}$ & $8.64\cdot10^{-6}$ & $0.076$ \\
$^{9}$Be & $1.36\cdot10^{7}$ & $1.35\cdot10^{-40}$ & $4.27\cdot10^{-5}$ & $1.000$ \\
$^{12}$C & $1.34\cdot10^{7}$ & $3.46\cdot10^{-40}$ & $5.05\cdot10^{-4}$ & $0.989$ \\
$^{13}$C & $1.34\cdot10^{7}$ & $3.93\cdot10^{-40}$ & $5.05\cdot10^{-4}$ & $0.011$ \\
$^{21}$Ne & $5.68\cdot10^{6}$ & $3.11\cdot10^{-41}$ & $6.93\cdot10^{-3}$ & $2.70\cdot10^{-3}$ \\
$^{23}$Na & $4.38\cdot10^{6}$ & $9.86\cdot10^{-42}$ & $1.08\cdot10^{-2}$ & $1.000$ \\
$^{25}$Mg & $3.84\cdot10^{6}$ & $9.61\cdot10^{-42}$ & $1.64\cdot10^{-2}$ & $0.100$ \\
$^{33}$S & $2.51\cdot10^{5}$ & $7.70\cdot10^{-42}$ & $6.57\cdot10^{-2}$ & $7.60\cdot10^{-3}$ \\
$^{34}$S & $5.38\cdot10^{6}$ & $5.34\cdot10^{-42}$ & $6.57\cdot10^{-2}$ & $0.043$ \\
$^{35}$Cl & $1.70\cdot10^{5}$ & $7.35\cdot10^{-42}$ & $8.73\cdot10^{-2}$ & $0.758$ \\
$^{64}$Ni & $7.31\cdot10^{6}$ & $1.93\cdot10^{-41}$ & $0.85$ & $9.26\cdot10^{-3}$ \\
$^{69}$Ga & $9.19\cdot10^{5}$ & $1.18\cdot10^{-41}$ & $1.33$ & $0.601$ \\
$^{98}$Mo & $4.60\cdot10^{6}$ & $8.78\cdot10^{-42}$ & $5.15$ & $0.241$ \\
$^{100}$Ru & $3.22\cdot10^{6}$ & $7.22\cdot10^{-42}$ & $6.32$ & $0.126$ \\
$^{104}$Pd & $2.46\cdot10^{6}$ & $8.02\cdot10^{-42}$ & $7.70$ & $0.111$ \\
$^{106}$Cd & $2.21\cdot10^{5}$ & $1.09\cdot10^{-41}$ & $9.30$ & $0.013$ \\
$^{108}$Cd & $1.68\cdot10^{6}$ & $1.08\cdot10^{-41}$ & $9.30$ & $8.90\cdot10^{-3}$ \\
$^{110}$Cd & $2.92\cdot10^{6}$ & $6.12\cdot10^{-42}$ & $9.30$ & $0.125$ \\
$^{112}$Sn & $6.93\cdot10^{5}$ & $2.25\cdot10^{-41}$ & $11.15$ & $9.70\cdot10^{-3}$ \\
$^{114}$Sn & $2.02\cdot10^{6}$ & $9.34\cdot10^{-42}$ & $11.15$ & $6.60\cdot10^{-3}$ \\
$^{116}$Sn & $3.31\cdot10^{6}$ & $5.72\cdot10^{-42}$ & $11.15$ & $0.145$ \\
\noalign{\smallskip}\hline
\end{tabular}
\end{center}
\label{tab:crs}
\end{table}

\section{Bound-beta beams}
\label{bbbeam}

A possible source of monochromatic antineutrinos 
are radioactive nuclei that undergo bound-beta decay,
a process confirmed experimentally in 1992~\cite{Jung:1992pw}.
We consider fully ionized atoms, where the bound-beta decay fraction is largest due to the 
two available vacancies in the $1s$ orbital.
In general, the $Q$ value of the bound-beta decay will not exactly match the $Q$ value
at the target, even if the reactions involve the same nuclei, 
because of nuclear recoil and differences in electron binding energies.

By accelerating the ionized atoms and storing them in a storage ring with straight sections,
the $Q$ value of any antineutrino source through bound-beta decay, $Q_s$, can be matched to the 
$Q$ value in the macroscopic target at rest, $Q_t$, even for different reactions.

For bound-beta decay of a fully ionized atom we find:
\begin{equation}
Q_{b\beta,I}=m(^A_ZX)-m(^A_{Z+1}Y)+B_{tot}(Z)-B_{tot}(Z+1)+B_{n,I},
\end{equation}
where $B_{tot}$ is the total binding energy of all electrons of the atom~\cite{Btot}.
$B_{n,I}$ is the binding energy of a single electron in orbital $n$ in an otherwise fully ionized atom:
\begin{equation}
B_{n,I}=m_e\left(1-\sqrt{1-\left(\alpha\frac{Z+1}{n}\right)^2}\right).
\end{equation}
Since the $Q$ values involved in beta decay are much smaller than the nuclear mass, 
the largest part of the energy release in bound-beta decay goes into the 
kinetic energy of the antineutrino produced:
\begin{equation}
E_{\bar\nu s}=Q_s-\frac{Q_s^2}{2m}.
\end{equation}

\begin{figure}
\resizebox{0.50\textwidth}{!}{\includegraphics{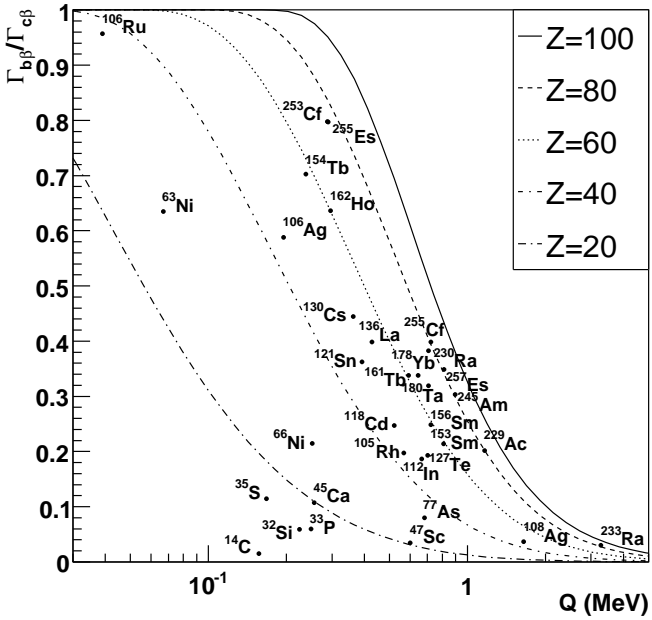}}
\caption{Fraction of beta decays of fully ionized atoms where the emitted electron is captured 
in an orbital.}
\label{fig:fb}
\end{figure}

Tab.~\ref{tab:fb} lists nuclei with $\Gamma_{b\beta}/\Gamma_{c\beta}\geq0.2$,
and the dependence on $Z$ and $Q$ is illustrated in Fig.~\ref{fig:fb}. 

\begin{table}
\caption{List of nuclei with a bound-beta decay fraction of 0.20 or more.}
\label{tab:fb}
\begin{center}
\begin{tabular}{cccc}
\hline\noalign{\smallskip}
   & $Q_{b\beta,I}$ &  $t_{1/2,I}$ &  $\Gamma_{b\beta}/\Gamma_{c\beta}$ \\
   &  (keV)        &    (s)        &        \\
\noalign{\smallskip}\hline\noalign{\smallskip}
$^{63}$Ni &   74.8 & $1.37\cdot10^{9}$ &  0.63 \\
$^{66}$Ni &  259.9 & $1.62\cdot10^{5}$ &  0.21 \\
$^{106}$Ru &   60.4 & $2.52\cdot10^{6}$ &  0.96 \\
$^{106}$Ag &  219.8 & $6.72\cdot10^{2}$ &  0.59 \\
$^{118}$Cd &  547.0 & $2.40\cdot10^{3}$ &  0.25 \\
$^{121}$Sn &  419.4 & $6.67\cdot10^{4}$ &  0.36 \\
$^{130}$Cs &  396.7 & $1.06\cdot10^{3}$ &  0.44 \\
$^{136}$La &  466.0 & $3.86\cdot10^{2}$ &  0.40 \\
$^{153}$Sm &  853.5 & $1.38\cdot10^{5}$ &  0.21 \\
$^{156}$Sm &  768.8 & $2.69\cdot10^{4}$ &  0.25 \\
$^{154}$Tb &  289.0 & $2.73\cdot10^{4}$ &  0.70 \\
$^{161}$Tb &  643.9 & $4.24\cdot10^{5}$ &  0.34 \\
$^{162}$Ho &  350.5 & $3.78\cdot10^{2}$ &  0.64 \\
$^{178}$Yb &  705.2 & $3.17\cdot10^{3}$ &  0.34 \\
$^{180}$Ta &  774.0 & $2.15\cdot10^{4}$ &  0.32 \\
$^{230}$Ra &  810.5 & $3.80\cdot10^{3}$ &  0.38 \\
$^{229}$Ac & 1268.8 & $3.20\cdot10^{3}$ &  0.20 \\
$^{245}$Am & 1018.7 & $5.63\cdot10^{3}$ &  0.30 \\
$^{253}$Cf &  420.7 & $4.13\cdot10^{5}$ &  0.80 \\
$^{255}$Cf &  854.4 & $3.45\cdot10^{3}$ &  0.40 \\
$^{255}$Es &  426.8 & $9.29\cdot10^{5}$ &  0.80 \\
$^{257}$Es &  947.8 & $4.82\cdot10^{5}$ &  0.35 \\
\noalign{\smallskip}\hline
\end{tabular}
\end{center}
\end{table}

The neutrino energy in the lab frame, $E_{\bar\nu}$, is related to the neutrino energy at the source, $E_{\bar\nu s}$,
coming from the decay of an atom with speed $\beta$ and boost $\gamma=(1-\beta^2)^{-1/2}$ by
\begin{equation}\label{eq:boost}
E_{\bar\nu}=E_{\bar\nu s}\gamma(1+\beta\cos\theta'),
\end{equation}
where $\theta'$ is the angle between the emitted neutrino and the direction of the atom in the frame of the decaying atom.
In the lab frame, the angle $\theta$ between the neutrino direction and the decaying nucleus is related to $\theta'$ as:
\begin{equation}
\cos\theta=\frac{\cos\theta'+\beta}{1+\beta\cos\theta'}.
\end{equation}

To obtain the resonant neutrino energy of the target in the direction of motion of the source atoms ($\theta=0$),
they need to be accelerated at a speed $\beta$:
\begin{equation}
\beta=\frac{E_{\bar\nu t}^2-E_{\bar\nu s}^2}{E_{\bar\nu t}^2+E_{\bar\nu s}^2}.
\end{equation}
The value of $\beta$ may be positive or negative: 
a positive value of $\beta$ corresponds to the target being positioned in the forward direction
of the accelerated ions,
a negative value of $\beta$ corresponds to the target being positioned in the backward direction
of the accelerated ions.

The neutrino energy in the lab frame depends on the decay angle, and monochromaticity is only achieved for 
a limited range of angles.
To produce neutrinos within a fraction $x$ of the resonance width,
the momentum of the beam is tuned such that 
the energy at $\theta=0$ is $E_{\bar\nu t}+x\Gamma$ and a target is used in the shape of 
cone with an angle $\theta$ where the incident neutrino energy is $E_{\bar\nu t}-x\Gamma$.
From Eq.~\ref{eq:boost} it follows that:
\begin{equation}
\cos\theta=1-\frac{2x\Gamma}{E_{\bar\nu s}|\beta|\gamma}.
\end{equation}

Cause of neutrino energy spread is also the momentum spread $\frac{\Delta p}{p}$ of the ions in the storage ring.
$\frac{\Delta E_{\bar\nu}}{E_{\bar\nu}}$,
the resulting energy spread of the neutrinos, 
derived from Eq.~\ref{eq:boost}, is
\begin{equation}
\frac{\Delta E_{\bar\nu}}{E_{\bar\nu}}=\frac{\Delta p}{p}|\beta|,
\end{equation}
which implies that for non-relativistic beams the energy spread of the neutrinos
is much smaller than the momentum spread of the decaying ions.
Requiring that the contribution from the momentum spread of the beam
be negligible compared to the radiative width of the target
corresponds to the following requirement on the ion beam:
\begin{equation}
\frac{\Delta p}{p}\ll
\left(\frac{\Delta p}{p}\right)_{max}=
\frac{\Gamma}{|\beta| E_\nu}.
\end{equation}


If the momentum spread of the beam is not negligible,
and we assume a Gaussian energy spread,
the corresponding reduction in effective peak cross-section can be calculated as:
\begin{equation}
\frac{\sigma_{eff}}{\sigma_{peak}}=
\sqrt\frac{\pi}{8}\frac{1}{y}\mbox{erfc}\left(\frac{1}{\sqrt{8}y}\right)e^{\frac{1}{8y^2}},
\end{equation}

where $y=\frac{\Delta E}{\Gamma}$.
This ratio for $\Delta E=\Gamma$ is 0.43 while it is 0.99 for $\Delta E=0.05\Gamma$.
Accordingly the number of events in the last column of Tab.~\ref{tab:expdem} and \ref{tab:exposc}
should be scaled by these factors.

\section{Interaction rates}
\label{rates}

We investigate possible experiments, where
ionized radioactive atoms are stored in a storage ring with
two straight sections, which produce the useful neutrinos, and two curved sections
to allow continuous circulation of the atoms.
Two targets  are placed on the axis of the straight sections.
To see whether these experiments are feasible, 
we make a set of simplifying assumptions:
\begin{itemize}
\item The length of the curved sections is neglected with respect to the length of the straight sections.
\item The size of the storage ring is taken to be negligibly small with respect to the targets.
This results in a point source with neutrinos emitted in both directions.
\item The targets are presumed to be 100\% pure isotope.
Several targets that we find promising have a natural abundance above 50\%, 
others have natural abundance below 1\% and a costly enrichment may be needed to 
reach the interaction rate quoted here.
\item The atoms in the storage ring are presumed to have zero momentum spread.
\end{itemize}

We consider two types of experiments:
\begin{itemize}
\item A \emph{demonstration} experiment: two targets of mass $m=10^3$\,kg are placed as close as possible to the source.
The targets have the shape of a cone with an opening angle of $\theta$,
corresponding to the zone of monochromaticity. 
The length of the targets depends on the opening angle and the density $\rho$ of the element:
$L=\left(\frac{3m}{\pi\theta_{max}^2\rho}\right)^{1/3}$.
\item An \emph{oscillation} experiment: two targets of  mass $m=10^3$\,kg shaped as truncated cones
with a length equal to the distance between an oscillation maximum and an oscillation minimum $L=\frac{2\pi E_\nu}{\Delta m_{32}^2}$
are positioned at a distance of half the target length.
We use $\Delta m_{32}^2=2.4\cdot10^{-3}$\,eV$^2$~\cite{Adamson:2008zt}. 
The available volumes are limited by the oscillation length:
$V=\frac{13\pi}{12}L_{osc}^3\theta^2$.
If the volumes are too small to contain the masses, we reduce the target mass. 
If the volumes are larger than needed to contain the target, 
we reduce $\theta$ to increase the average cross-section over the target volume.
\end{itemize}

\begin{figure}
\resizebox{0.50\textwidth}{!}{\includegraphics{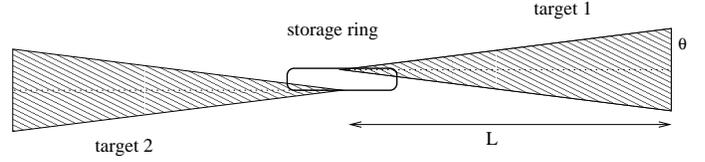}}
\caption{Considered setup for an experiment to demonstrate antineutrino induced electron capture.}
\label{fig:setup_dem}
\end{figure}

\begin{figure}
\resizebox{0.50\textwidth}{!}{\includegraphics{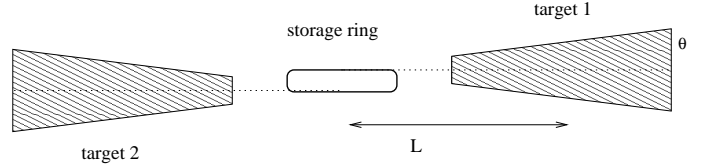}}
\caption{Considered setup for an experiment to study antineutrino induced electron capture 
at the first neutrino oscillation maximum.}
\label{fig:setup_osc}
\end{figure}

The reaction rate $N_A$, defined as the number of neutrino interactions per decaying atom, 
can be expressed as
\begin{equation}
N_A=\frac{\Gamma_{b\beta}}{\Gamma}\frac{1-\cos\theta'}{2}\varepsilon_{BW}\sigma_{peak} L \frac{\rho}{m},
\end{equation}
where $\varepsilon_{BW}$ is the ratio of the average cross-section and the peak cross-section:
\begin{equation}
\varepsilon_{BW}=\frac{1}{2x\Gamma}\int_{Q-x\Gamma}^{Q+x\Gamma}dE\frac{\Gamma^2/4}{(E-Q)^2+\Gamma^2/4}
=\frac{\arctan(2x)}{2x}.
\end{equation}
The factor $\frac{1-\cos\theta'}{2}$ corresponds to the fraction 
of neutrinos that traverses the target.

Apart from the interaction rate per decaying nucleus, we also put requirements on the 
number of interactions per unit of time per stored atom:
\begin{equation}
N_T=\frac{N_A}{\tau},
\end{equation}
where $\tau$ is the average lifetime of the ionized atom.
As an additional constraint, we consider the number of interactions per unit of energy
needed to ionize and accelerate the atom: 
\begin{equation}
N_E=\frac{N_A}{B_{tot}+m(\gamma-1)}.
\end{equation}


Tab.~\ref{tab:expdem} lists the demonstration experiments that
have $N_A>10^{-20}$, $N_T>10^{-20}$\,y$^{-1}$ and $N_E>10^{-10}$\,J$^{-1}$.
The most favourable case, using a source of $^{253}$Cf and a target of $^{121}$Sb,
corresponds to a rate of one interaction for $8.3\cdot10^{18}$ decaying atoms.
Tab.~\ref{tab:exposc} lists the oscillation experiments that
have  $N_A>10^{-22}$, $N_T>10^{-22}$\,y$^{-1}$ and $N_E>10^{-12}$\,J$^{-1}$.
The most favourable case, using a source of $^{35}$S and a target of $^{106}$Cd,
corresponds to a rate of one interaction for $3.2\cdot10^{21}$ decaying atoms.

The rates we find compare favorably with the rates for other 
low-energy neutrino reactions.
For example, antineutrino-proton scattering using reactor neutrinos at the first atmospheric oscillation maximum 
of 3\,MeV anti-neutrinos requires
$8\cdot10^{23}$ fissions per interaction for a target of $10^3$\,kg of pure hydrogen~\cite{Declais:1994ma},
more than two orders of magnitude above the best configuration of Tab.~\ref{tab:exposc}.
However, this advantage in rate is lost when one compares the flux:
a typical nuclear reactor produces $\approx10^{20}$ fissions per second, 
while beams of future projects for beams of radioactive atoms
are $\approx2\cdot10^{13}$ decays per second~\cite{Donini:2006tt}.

Another major challenge is the requirement on the beam monochromaticity.
Tab.~\ref{tab:expdem} and \ref{tab:exposc} indicate that a momentum spread
of $10^{-1}$ to $10^{-7}$ is required for a demonstration experiment,
while oscillation experiments require a beam monochromaticity of at least to $10^{-7}$.
Beams of ordered heavy ions have been show to reach momentum spreads
below $5\cdot10^{-7}$~\cite{Steck:1996me},
and a transverse beam size of a few $\mu$m,~\cite{Steck:2003},
but only at very low intensities of a few thousand stored ions.
Three-dimensionally crystallized beams~\cite{Ikegami:2006kt}
could allow for higher intensity
beams, but they have not yet been demonstrated experimentally.

\begin{table*}
\caption{List of demonstration experiments with a yield of at least $10^{-20}$ interactions per
decaying atom, $10^{-20}$ interactions per stored atom per year and $10^{-10}$ events per Joule.}
\begin{center}
\begin{tabular}{ccccccccccc}
\hline\noalign{\smallskip}
source&$\Gamma_{b\beta}/\Gamma$& $\beta$   & $\tau$  &target&natural      &$Q_t$&$\theta$  &$L$&$(\Delta p/p)_{max}$&$N_A$   \\
      &                        &$(10^{-3})$&  (s)    &      &abundance(\%)&(keV)&(mrad)&(m)&            &$(10^{-22})$    \\
\noalign{\smallskip}\hline\noalign{\smallskip}
$^{77}$As & 0.080 & $-2.06$ & $1.90\cdot10^{5}$ & $^{112}$Sn$_{1s}$ &  1.0 &  693 & 177 &   1.61 &$7.81\cdot10^{-3}$ &  486 \\
$^{108}$Ag & 0.035 & $ 0.45$ & $2.01\cdot10^{2}$ & $^{108}$Cd$_{1s}$ &  0.9 & 1676 & 222 &   1.31 &$1.23\cdot10^{-2}$ &  164 \\
$^{112}$In & 0.082 & $ 1.10$ & $1.10\cdot10^{3}$ & $^{112}$Sn$_{1s}$ &  1.0 &  693 & 242 &   1.31 &$1.46\cdot10^{-2}$ &  764 \\
$^{121}$Sn & 0.362 & $ 1.85$ & $9.62\cdot10^{4}$ & $^{121}$Sb$_{1s}$ & 57.2 &  420 & 251 &   1.31 &$1.56\cdot10^{-2}$ &  390 \\
$^{127}$Te & 0.193 & $-56.00$ & $4.11\cdot10^{4}$ & $^{112}$Sn$_{1s}$ &  1.0 &  693 &  35 &   4.77 &$2.87\cdot10^{-4}$ &  121 \\
$^{161}$Tb & 0.338 & $73.35$ & $6.14\cdot10^{5}$ & $^{112}$Sn$_{1s}$ &  1.0 &  693 &  29 &   5.45 &$2.19\cdot10^{-4}$ &  211 \\
$^{161}$Tb & 0.338 & $38.66$ & $6.13\cdot10^{5}$ & $^{112}$Sn$_{2s}$ &  1.0 &  669 &   6 &  15.11 &$9.91\cdot10^{-6}$ &  139 \\
$^{161}$Tb & 0.338 & $33.55$ & $6.13\cdot10^{5}$ & $^{112}$Sn$_{3s}$ &  1.0 &  666 &   1 &  42.70 &$4.37\cdot10^{-7}$ &  134 \\
$^{178}$Yb & 0.338 & $-17.53$ & $4.57\cdot10^{3}$ & $^{112}$Sn$_{1s}$ &  1.0 &  693 &  61 &   3.28 &$9.18\cdot10^{-4}$ &  486 \\
$^{178}$Yb & 0.338 & $-52.28$ & $4.58\cdot10^{3}$ & $^{112}$Sn$_{2s}$ &  1.0 &  669 &   6 &  16.22 &$7.33\cdot10^{-6}$ &  101 \\
$^{245}$Am & 0.303 & $-0.20$ & $8.13\cdot10^{3}$ & $^{164}$Er$_{1s}$ &  1.6 & 1019 & 961 &   0.49 &$2.14\cdot10^{-1}$ &  715 \\
$^{253}$Cf & 0.798 & $-1.10$ & $5.95\cdot10^{5}$ & $^{121}$Sb$_{1s}$ & 57.2 &  420 & 326 &   1.10 &$2.62\cdot10^{-2}$ & 1206 \\
$^{255}$Es & 0.733 & $-15.64$ & $1.34\cdot10^{6}$ & $^{121}$Sb$_{1s}$ & 57.2 &  420 &  87 &   2.67 &$1.85\cdot10^{-3}$ &  186 \\
\noalign{\smallskip}\hline
\end{tabular}
\end{center}
\label{tab:expdem}
\end{table*}

\begin{table*}
\caption{List of oscillation experiments with a yield of at least $10^{-22}$ interactions per
decaying atom, $10^{-22}$ interactions per stored atom per year and $10^{-12}$ events per Joule.}
\begin{center}
\begin{tabular}{ccccccccccc}
\hline\noalign{\smallskip}
source&$\Gamma_{b\beta}/\Gamma$& $\beta$   & $\tau$  &target&natural      &$Q_t$&$\theta$  &$L$&$(\Delta p/p)_{max}$&$N_A$   \\
      &                        &$(10^{-3})$&  (s)    &      &abundance(\%)&(keV)&($\mu$rad)&(m)&            &$(10^{-24})$    \\
\noalign{\smallskip}\hline\noalign{\smallskip}
$^{35}$S & 0.114 & $-12.0$ & $9.99\cdot10^{6}$ & $^{ 35}$Cl$_{2s}$ & 75.8 &  167 & 540 &     86 &$3.79\cdot10^{-7}$ &  190 \\
$^{35}$S & 0.114 & $143.2$ & $1.01\cdot10^{7}$ & $^{106}$Cd$_{3s}$ &  1.2 &  196 & 181 &    101 &$2.63\cdot10^{-7}$ &  315 \\
$^{66}$Ni & 0.215 & $-44.2$ & $2.34\cdot10^{5}$ & $^{ 33}$S$_{2s}$ &  0.8 &  249 & 259 &    128 &$4.64\cdot10^{-8}$ &  174 \\
$^{153}$Nd & 0.014 & $ 38.0$ & $4.56\cdot10^{1}$ & $^{  6}$Li$_{1s}$ &  7.6 & 3508 &  10 &   1813 &$6.48\cdot10^{-11}$ &  115 \\
$^{152}$Pm & 0.013 & $-11.8$ & $3.57\cdot10^{2}$ & $^{  6}$Li$_{1s}$ &  7.6 & 3508 &  10 &   1813 &$2.09\cdot10^{-10}$ &  110 \\
$^{155}$Pm & 0.015 & $ 70.1$ & $6.00\cdot10^{1}$ & $^{  6}$Li$_{1s}$ &  7.6 & 3508 &  10 &   1813 &$3.51\cdot10^{-11}$ &  108 \\
$^{153}$Sm & 0.214 & $ 63.9$ & $1.99\cdot10^{5}$ & $^{ 69}$Ga$_{3s}$ & 60.1 &  910 &  22 &    470 &$4.80\cdot10^{-9}$ &  148 \\
$^{161}$Tb & 0.338 & $ 33.6$ & $6.13\cdot10^{5}$ & $^{112}$Sn$_{3s}$ &  1.0 &  666 &  32 &    344 &$4.37\cdot10^{-7}$ &  115 \\
$^{169}$Dy & 0.019 & $ 74.9$ & $5.63\cdot10^{1}$ & $^{  6}$Li$_{1s}$ &  7.6 & 3508 &  10 &   1813 &$3.29\cdot10^{-11}$ &  127 \\
$^{171}$Ho & 0.019 & $ 73.8$ & $7.65\cdot10^{1}$ & $^{  6}$Li$_{1s}$ &  7.6 & 3508 &  10 &   1813 &$3.34\cdot10^{-11}$ &  132 \\
$^{230}$Ra & 0.383 & $115.3$ & $5.52\cdot10^{3}$ & $^{ 69}$Ga$_{3s}$ & 60.1 &  910 &  22 &    470 &$2.66\cdot10^{-9}$ &  292 \\
$^{233}$Ra & 0.030 & $ 39.8$ & $4.31\cdot10^{1}$ & $^{  6}$Li$_{1s}$ &  7.6 & 3508 &  10 &   1813 &$6.19\cdot10^{-11}$ &  252 \\
$^{232}$Ac & 0.024 & $-81.2$ & $1.72\cdot10^{2}$ & $^{  6}$Li$_{1s}$ &  7.6 & 3508 &  10 &   1813 &$3.03\cdot10^{-11}$ &  123 \\
$^{255}$Cf & 0.398 & $ 62.9$ & $4.99\cdot10^{3}$ & $^{ 69}$Ga$_{3s}$ & 60.1 &  910 &  22 &    470 &$4.87\cdot10^{-9}$ &  274 \\
$^{257}$Es & 0.349 & $-40.8$ & $6.97\cdot10^{5}$ & $^{ 69}$Ga$_{3s}$ & 60.1 &  910 &  22 &    470 &$7.52\cdot10^{-9}$ &  195 \\
\noalign{\smallskip}\hline
\end{tabular}
\end{center}
\label{tab:exposc}
\end{table*}

\appendix

\section{Bound-beta decay}\label{sec:appendix}
\label{bbdecay}

%
%
%

The probability of bound-beta decay depends on the number of free orbits in 
the decaying nucleus, and thus on the level of ionization.
The ratio of bound-beta to continuous beta decay 
depends on the overlap of the electron wave function and the nucleus~\cite{bahcall},
and is non-negligible only for the $s$ orbitals with zero angular momentum.
For high-$Z$ nuclei, relativistic effects have a sizable effect on the shapes of the orbitals,
and we use the relation~\cite{mikaelyan}:
\begin{equation}\label{eq:fbb}
\frac{\Gamma_{b\beta}}{\Gamma_{c\beta}}=
n_f\pi\left(\frac{\alpha}{n}\right)^3
\frac{(Z+1)^{2.87+6.2\cdot10^{-3}(Z+1)}}
{f(Z+1,Q_{c\beta})}
\left(\frac{Q_{b\beta}}{m_e}\right)^2,
\end{equation}
where $\alpha$ is the fine structure constant,
$n_f$ is the number of free orbitals at a given energy,
$n$ is the quantum number of energy,
$Q_{b\beta}$ is the energy release in a bound-beta decay,
$Q_{c\beta}$ is the energy release in a continuous-beta decay, and
$m_e$ the mass of the electron.
We calculate the Fermi integral for continuous beta decay $f(Z,Q)$
 relativistically and include effects of Coulomb screening:
\begin{equation}
f(Z,Q)=\frac{1}{m_e^5}\int(Q-T)^2(T+m_e)\sqrt{T^2+m_eT}F(Z,T)dT,
\end{equation}
where $F(Z,T)$ is the Coulomb screening function, depending on the atomic number $Z$ and the 
kinetic energy of the electron, $T$.
We use the relativistic Fermi function
\begin{equation}
F(Z,T)=\frac{2+2S}{((2S)!)^2}(2pR)^{2S-2}e^{\pi\eta}|(S-1+i\eta)!|^2,
\end{equation}
where $p$ is the electron momentum, 
$S=\sqrt{1-\alpha^2Z^2}$, $\eta=Z\alpha\beta^{-1}$, and $R$ the nuclear radius, approximated by
$R=1.2\,{\rm fm}/(\hbar c)A^{1/3}$.

\end{document}